%% file: main.tex
\documentclass{article}
\usepackage{spconf,amsmath,graphicx}
\usepackage{times}  
\usepackage{helvet}  
\usepackage{courier}  
\usepackage{url}  
\usepackage{booktabs}
\usepackage{subcaption}
\usepackage{xcolor}
\usepackage{color}
\usepackage{todonotes}
\usepackage{multirow}

\title{ASR ERROR CORRECTION and DOMAIN ADAPTATION USING MACHINE TRANSLATION}
\name{Anirudh Mani$^{1,*}$, Shruti Palaskar$^{2,}$\sthanks{Equal contribution}, Nimshi Venkat Meripo$^{1}$, Sandeep Konam$^{1}$ and Florian Metze$^{1,2}$}
\address{
  $^{1}$ Abridge AI, $^{2}$ Carnegie Mellon University\\ 
  \small{\texttt{\{amani, venkatm, san\}@abridge.com}}\\
  \small{\texttt{\{spalaska, fmetze\}@cs.cmu.edu}}\\
}
\begin{document}
%
\maketitle
\begin{abstract}

\input{sections/abstract.tex}
\end{abstract}


\section{Introduction}
\label{sec:intro}
\input{sections/introduction.tex}

\section{Related Work}
\label{sec:related_work}
\input{sections/related_work.tex}

\section{Domain Adaptation for Error Correction}
\label{sec:domain_adapt}
\input{sections/method.tex}

\section{Experimental Setup}
\label{sec:experimental_setup}
\input{sections/experimental_setup.tex}

\section{Results and Discussions}
\label{sec:results_discussions}
\input{sections/results_discussion.tex}

\section{Conclusion}
\label{sec:conclusion}
\input{sections/conclusion.tex}

\small 
\section{Acknowledgements}
\label{sec:acknowledgement}
We thank Prof. Alan Black, Sai Prabhakar Pandi Selvaraj and Ben Schloss for many helpful discussions and valuable feedback. We also thank the University of Pittsburgh Medical Center (UPMC) and Abridge AI Inc. for providing access to de-identified data; Ben Schloss, Steven Coleman, and Deborah Osakue for data business development and annotation management.

Shruti Palaskar received funding via Amazon Web Services (AWS) Machine Learning Research Awards and the Center for Machine Learning and Health (CMLH) at Carnegie Mellon University through the Pittsburgh Health Data Alliance. This work used the Extreme Science and Engineering Discovery Environment (XSEDE), which is supported by National Science Foundation grant number ACI-1548562. Specifically, it used the Bridges system, which is supported by NSF award at the Pittsburgh Supercomputing Center (PSC).




\bibliographystyle{IEEEbib}
\bibliography{strings,refs}

\end{document}

%% file: sections/abstract.tex
Off-the-shelf pre-trained Automatic Speech Recognition (ASR) systems are an increasingly viable service for companies of any size building speech-based products. While these ASR systems are trained on large amounts of data, domain mismatch is still an issue for many such parties that want to use this service as-is leading to not so optimal results for their task. We propose a simple technique to perform domain adaptation for ASR error correction via machine translation. The machine translation model is a strong candidate to learn a mapping from out-of-domain ASR errors to in-domain terms in the corresponding reference files. We use two off-the-shelf ASR systems in this work: Google ASR (commercial) and the ASPIRE model (open-source). We observe 7\% absolute improvement in word error rate and 4 point absolute improvement in BLEU score in Google ASR output via our proposed method. We also evaluate ASR error correction via a downstream task of Speaker Diarization that captures speaker style, syntax, structure and semantic improvements we obtain via ASR correction.


%% file: sections/introduction.tex
Cloud-based ASR systems are easily available to companies building speech-based products. These products cover a wide-range of use cases like speech transcriptions, language understanding, spoken language translation, information extraction, and summarization. Most of these use-cases involve transcribing speech and then performing various downstream language-processing tasks. In these scenarios, there is a 
break of domain in two places, one for speech-to-text where pre-trained ASR is trained on different domains of data, and another while optimizing NLP downstream tasks with transcriptions from pre-existing ASR trained on another domain. This is a break that also stems from being unable to train in-house competitive ASR on in-domain data alone, which has a lesser chance of out performing pre-trained ASRs on much larger data, even if it is out-of-domain. Towards solving this problem, we propose to carry out \textit{ASR error correction} via domain adaptation on two pre-existing ASRs: ASPIRE model \cite{peddinti2015jhu} which is an open-source resource trained on conversational, broadcast, and read speech, and Google Speech API\footnote{https://cloud.google.com/speech-to-text/} which is trained on large quantities of English speech. 

We propose to learn an adaptation module that goes from hypothesis of pre-trained ASR towards reference text, in the process, grounding ASR hypothesis to the domain of the data, and learning to fix any systematic errors the pre-trained ASR makes due to domain mismatch. We evaluate the benefits of this adaptation module in terms of ASR output correction or spelling correction i.e. at a syntactic level, semantic level, on a separate speaker diarization model and on word error rate. Improving ASR transcriptions will also improve reader experience and downstream language processing tasks in addition to addressing the domain mismatch problem.

\begin{table}[t]
  \centering
  \begin{tabular}{p{2cm}p{5.5cm}}
    \toprule
    Model & Transcript 	 \\
    \midrule
    Reference 	& that's why you're on the \textit{coumadin} \\
    Google ASR		& that 's why you 're on the \textbf{cool midi} \\ \midrule
    Reference & but the \textit{coumadin} stays there for days \\
    Google ASR & but the \textbf{cool molina} stays there for days \\ \midrule
    Reference & we can use \textit{coumadin} the same way \\
    Google ASR & we can use \textbf{cumin in} the same way . \\
    \bottomrule
  \end{tabular}
  \caption{Examples from Reference and Google ASR transcription for a particular medical word \textit{``Coumadin''}. We observe the same medical word mis-transcribed in many different ways. In this work, we investigate whether adapting transcription to domain and context can help reduce such errors.}
  \label{tab:coumadin_example}
\end{table}

Using the reference texts and pre-trained ASR hypothesis, we have access to data that is in-domain (reference text) and out-of-domain (hypothesis from ASR), both of which are transcriptions of the same speech signal. To learn an adaptation from out-of-domain to in-domain data, we can model the task as an automatic machine translation problem where the translation-module learns to adapt domains by learning differences in style of the data syntactically, semantically and structurally. Using this, we aim to automatically correct any systematic speech recognition errors due to domain mismatch.

While ASR error correction hopes to improve transcription quality, in addition to evaluating in terms of Word Error Rate (WER) at utterance levels, we also evaluate domain-specific errors, medical terms in our case, via a medical WER metric. Additionally, transcription improvement can also be measured using a downstream task of speaker diarization or classification. If the stylistic characteristics of transcription improve, an improvement in the speaker classification task is also expected. In this work, we also evaluate ASR error correction on this task to show its utility.

%% file: sections/related_work.tex
While the need for ASR correction has become more and more prevalent in recent years with the successes of large-scale ASR systems, machine translation and domain adaptation for error correction are still relatively unexplored. D'Haro and Banchs \cite{d2016automatic} first explored the use of machine translation to improve automatic transcription and they applied it to robot commands dataset and human-human recordings of tourism queries dataset. ASR error correction has also been performed based on ontology-based learning in \cite{anantaram2018repairing}. They investigate the use of including accent of speaker and environmental conditions on the output of pre-trained ASR systems. Their proposed approach centers around bio-inspired artificial development for ASR error correction. \cite{shivakumar2019learning} explore the use of noisy-clean phrase context modeling to improve ASR errors. They try to correct unrecoverable errors due to system pruning from acoustic, language and pronunciation models to restore longer contexts by modeling ASR as a phrase-based noisy transformation channel. Domain adaptation with off-the-shelf ASR has been tried for pure speech recognition tasks in high and low resource scenarios with various training strategies \cite{swietojanski2014learning,swietojanski2015differentiable,meng2017unsupervised,sun2017unsupervised,shinohara2016adversarial,dalmia2018domain} but the goal of these models was to build better ASR systems that are robust to domain change. \cite{selvaraj2019medication} explore medical regimen extraction on the corpus we use in this work, which is another application where domain adaptation for ASR transcription can help improve the quality of domain-specific downstream tasks.

%% file: sections/method.tex
Using the reference texts and pre-trained ASR hypothesis, we have access to parallel data that is in-domain (reference text) and out-of-domain (hypothesis from ASR), both of which are transcriptions of the same speech signal. With this parallel data, we now frame the adaptation task as a translation problem, and also describe the speaker diarization model.

\subsection{Machine Translation Models}
\label{ssec:models_mt}

\paragraph*{Sequence-to-Sequence Models}: Sequence-to-sequence (S2S) models \cite{Sutskever2014} have been applied to various sequence learning tasks including speech recognition and machine translation. Attention mechanism \cite{bahdanau2014neural} is used to align the input with the output sequences in these models. The encoder is a deep stacked Recurrent Neural Network (RNN) and the decoder is usually a shallower uni-directional RNN acting as a language model for decoding the input sequence into either the transcription (ASR) or the translation (MT). Attention-based S2S models do not require alignment information between the source and target data, hence useful for monotonic and non-monotonic sequence-mapping tasks. In our work, we are mapping ASR output to reference hence it is a monotonic mapping task where we use this model.

\paragraph*{Transformers}: Transformers \cite{vaswani2017attention} are another common model architecture for machine translation that performed better than S2S models for certain datasets. Unlike S2S models, the Transformer model uses self-attention layers instead of RNNs to model varying length sequences. 
Self-attention layers also have a better capacity to learn longer range dependencies which is a challenge for RNNs. This makes this model another ideal candidate for us to try for this task.

\subsection{Speaker Diarization Model}
\label{ssec:models_srl}



We use a hierarchical bi-directional LSTM (BLSTM) \cite{hochreiter1997long} in an S2S type architecture to output a speaker role label, either $Doctor$ or $Patient$, for each utterance. Similar to machine translation in this work, each utterance is treated independently without any context from surrounding utterances for the speaker diarization model as well.

In the model architecture, first we use ELMo \cite{elmo} as a contextual embedder to encode words with their context in each utterance. 
The encoder which is a combination of a BLSTM and word-level attention mechanism, encodes all the words in an utterance $U_n$, into a fixed-length representation $u_n$, as:
$$h_{nm} = BLSTM(e_{nm}) $$
$$ a_{nm} =  softmax(W_{a} h_{nm} + b_a) \ \forall m=1...M $$
$$u_n =  \sum_{m}a_{nm} h_{nm} \ \ \forall n=1..b  $$

The decoder is a BLSTM layer followed by a sigmoid layer (operating on the BLSTM's hidden states).
Utterance representation from the encoder is then used by the decoder to provide utterance level context when classifying each utterance.
It produces the classification probability $\hat{s_{n}}$, corresponding to an utterance $U_n$, using it's representation $u_n$ from the encoder as:
$\hat{s_{n}} = \sigma{(W_c h_n + b_c)} \ \ \forall n=1...b, \ $ where$ \ [h_1...h_b] = BLSTM([u_1...u_b])$.
To train the network we use a binary cross-entropy loss function, given by 
$L_n = - (s_n log(\hat{s_{n}})+(1 - s_n) log(1 -\hat{s_{n}}))$.


%% file: sections/experimental_setup.tex
\subsection{Dataset}\label{subsec:Dataset}

We use a dataset of 3807 de-identified Doctor-Patient conversations containing 288,475 utterances split randomly into 230,781 training utterances and 28,847 for validation and test each. 
The total vocabulary for the machine translation task is 12,934 words in the ASR output generated using Google API and ground truth files annotated by humans in the training set. We only train word-based translation models in this study to match ASR transcriptions and ground truth with further downstream evaluations. There are 70.9\% of utterances with the Doctor label and remaining are Patient labels in the train set. Speaker diarization labels are human annotated and utterance level diarization is obtained using alignment across different ASR outputs and reference text (details below).

To choose domain-specific medical words, we use a pre-defined ontology by Unified Medical Language System (UMLS) \cite{umls}, giving us an exhaustive list of over 20,000 medications. First, each alternative name for the same medication in the list is regarded as a new medication name. Second, we keep only one word medications and remove alternate names. 
Furthermore, we filter out one letter entries, and medications with just a single or no occurrence in our ground truth test set. This results in a list of 200 most commonly occurring medications in the ground truth test set.


\paragraph*{Alignment:} Since the ground truth is at utterance level, and ASR system output transcripts are at word level, specific alignment handling techniques are required to match the output of multiple ASR systems. This is achieved using utterance level timing information i.e., start and end time of an utterance, and obtaining the corresponding words in the ASR system output transcript based on word-level timing information (start and end time of each word). To make sure same utterance ID is used across all ASR outputs and the ground truth, we first process our primary ASR output transcripts from Google Cloud Speech API based on the ground truth and create random training, validation and test splits. For each ground truth utterance in these dataset splits, we also generate corresponding utterances from ASPIRE output transcripts similar to the process mentioned above. This results in two datasets corresponding to Google Cloud Speech and ASPIRE ASR models, where utterance IDs are conserved across datasets. However, this does lead to ASPIRE dataset having a lesser utterances as we process Google ASR outputs first in an effort maximize the size of our primary ASR model dataset.

\subsection{Pre-trained ASR}
We use the Google Cloud Speech API for Google ASR transcription and the JHU ASPIRE model \cite{peddinti2015jhu} as two off-the-shelf ASR systems in this work. Google Speech API is a commercial service that charges users per minute of speech transcribed, while the ASPIRE model is an open-source ASR model. We explore the trends we observe in both--a commercial API as well as an open-source model. 

\subsection{Evaluations}

With the ultimate aim being ASR error correction based on domain adaptation, we can evaluate across many transcription-based evaluations: word error rate at utterance level, medical word error rate capturing domain-specific changes, BLEU scores capturing syntactic structure of outputs, and speaker diarization evaluation capturing speaker style, vocabulary, semantics and structure. Speaker diarization results and Medical WER, which is essentially WER calculated for each of 200 medications as described in \ref{subsec:Dataset}, are evaluated on Precision (P), Recall (R) and F1 score.

%% file: sections/results_discussion.tex
\subsection{Transcription Quality}

\begin{table}[t]
\centering
\resizebox{0.8\columnwidth}{!}{%
\renewcommand\arraystretch{1.0}
\begin{tabular}{llcc}
\toprule
& Transcript          & \multicolumn{2}{c}{Metric}      \\
&                     &                     WER ($\Downarrow$) & BLEU ($\Uparrow$)  \\ \midrule
Google ASR\phantom{s}   & ASR output  \phantom{s} & 41.0 & 52.1 \\
                        & S2S Adapted        & 34.1     & 56.4 \\ 
ASPIRE                  & ASR output         & 35.8 & 54.3 \\
                        & S2S Adapted        & 34.5 & 55.8 \\
\bottomrule
\end{tabular}}
\caption{Results for adaptive training experiments with Google ASR and ASPIRE model. We compare absolute gains in WER and BLEU scores with un-adapted ASR output.} 
\label{tbl:res_adapt1}
\end{table}

In Table \ref{tbl:res_adapt1} we compare the system performances on Google ASR and ASPIRE model in terms of WER and BLEU score improvement using S2S models. We consistently see significant gains in both the metrics by adapting the model. For Google ASR, the gains are more prominent with an absolute improvement of 7\% in WER and a 4 point absolute improvement in BLEU scores for the same model. On the same data, the Transformer model performs much badly, with a WER of 92.2 and BLEU score of 0.06, which we believe is due to less training data (similarly trained models used 21 times more data \cite{vaswani2017attention}). We observe smaller but significant gains on ASPIRE model output in comparison to Google ASR which might be due to no punctuation in the ASPIRE model output. While we could strip the Google ASR output of punctuation for a better comparison, it is an extra post-processing step and breaks the direct output modeling pipeline. If necessary, we can insert punctuation into ASPIRE model and our references as well.

In Table \ref{tbl:res_adapt_mwer}, we measure the transcript improvement specifically with respect to domain specific terms, in our case, medical words. 
We observe a high recall which shows that the model is able to generate medical words in the adapted model than it could in the ASR output alone. We see consistent improvements in Google ASR and ASPIRE model output. 


\begin{table}[t]
\centering
\resizebox{0.8\columnwidth}{!}{%
\renewcommand\arraystretch{1.0}
\begin{tabular}{llccc}
\toprule
& Transcript          & \multicolumn{3}{c}{\textbf{Medical WER}}      \\
&                     &                     P  & R & F1  \\ \midrule

Google ASR\phantom{s}   & ASR output  \phantom{s}    & 0.90  & 0.58  & 0.67   \\
                        & S2S Adapted        & 0.90  & 0.60  & 0.70 \\ \midrule
ASPIRE                  & ASR output        & 0.88 & 0.53 & 0.59\\
                        & S2S Adapted        &  0.86  & 0.54  & 0.60 \\

\bottomrule
\end{tabular}}
\caption{Results for adaptive training experiments on full data measured specifically on domain-specific words.}
\label{tbl:res_adapt_mwer}
\end{table}

\subsection{Speaker Diarization}

The purpose of using speaker diarization as a downstream task is to see if our machine translation model can learn to correct ASR errors based on speakers i.e., speaker style, speaker vocabulary, and syntactic, structural and semantic nuances. To this effect, we observe in Table \ref{tbl:srl} that our model learns to do exactly that. The results are higher for the \textit{Doctor} than for the \textit{Patient} class on the reference text itself which could be due to higher number of utterances spoken by the \textit{Doctor}. 

For \textit{Patient} utterances, we observe similar improvement trends for both the ASR model outputs. Overall, we get a 9\% percent improvement in Recall, 6\% increase in F1 score and 1\% drop in Precision in the Google ASR outputs, where as for the ASPIRE model outputs we get a 7\% percent improvement in Recall and an 4\% overall improvement in F1 score even though Precision drops by 2 \%.


\begin{table}[t]
\centering
\resizebox{0.95\columnwidth}{!}{%
\renewcommand\arraystretch{1.0}
\begin{tabular}{lccc}
\toprule
Model/Transcript    & \multicolumn{3}{c}{Diarization Metrics (Patient, Doctor)}      \\
                    &  P & R & F1  \\ \midrule
Reference Text      & 0.68, 0.83 & 0.48, 0.92 & 0.56, 0.87 \\ \midrule
Google ASR          & 0.75, 0.82 & 0.42, 0.95 & 0.54, 0.88 \\
Google ASR Adapted  & 0.74, 0.82 & 0.51, 0.92 & 0.60, 0.87 \\ \midrule 
ASPIRE ASR          & 0.69, 0.85 & 0.54, 0.91 & 0.60, 0.88 \\
ASPIRE ASR Adapted  & 0.67, 0.87 & 0.61, 0.89 & 0.64, 0.88 \\
\bottomrule
\end{tabular}}
\caption{Results for speaker diarization.}
\label{tbl:srl}
\end{table}



\subsection{Qualitative Analysis}



We look at the top 3 most frequent and least frequent medical terms in Table \ref{tbl:qual_mwer} and compute Medical WER on them in the ASR output and with the S2S adapted model to look closer at the F1 scores for domain-specific terms. More the number of medical term occurrences, more the model learns to predict them that leads to higher improvements in Recall and F1 scores. In the case of \textit{Coumadin}, we are able to recover from the different ASR mistakes shown in Table \ref{tab:coumadin_example}.

\begin{table}[t]
\centering
\resizebox{0.85\columnwidth}{!}{%
\renewcommand\arraystretch{1.0}
\begin{tabular}{lccc}
\toprule
Medical Word    & \multicolumn{3}{c}{Medical WER (ASR o/p, S2S adpt)}      \\
                    &  P & R & F1  \\ \midrule
                    \multicolumn{4}{c}{Most Frequent} \\ \midrule
Coumadin        & 0.98, 0.91  & 0.47, 0.73 & 0.64, 0.81 \\ 
Statin          & 0.98, 0.89 & 0.47, 0.63 & 0.63, 0.74 \\
Lisinopril      & 0.97, 0.87 & 0.38, 0.62 & 0.55, 0.72 \\ \midrule 
                    \multicolumn{4}{c}{Least Frequent} \\ \midrule
RID             & 0.88, 0.92 & 0.81, 0.78 & 0.84, 0.85 \\
Vitamin         & 0.97, 0.98 & 0.82, 0.81 & 0.89, 0.88 \\
ICAR            & 0.91, 0.92 & 0.69, 0.67 & 0.79, 0.78 \\
\bottomrule
\end{tabular}}
\caption{Qualitative analysis of Medical WER changes for 3 most frequent and least frequent medical words. Looking at the domain specific words highlights the utility of our model.}
\label{tbl:qual_mwer}
\end{table}

%% file: sections/conclusion.tex

We present a study to show that off-the-shelf ASR systems outputs can be optimized for specific domains as a post-processing step if we have access to the ASR hypothesis and reference texts via domain adaptation and machine translation. This method makes it viable for many domain specific applications and is easy to implement. We evaluate for ASR quality improvements: WER and BLEU scores, as well as specific improvements on domain specific terms like medical words in our case, and on a downstream task of speaker diarization that is directly affected by transcription quality. We see significant gains in the ASR transcription quality. Using the adaptation module, we can improve the generation of domain-specific words which the ASR mis-recognizes, and also improve speaker diarization. In the future, we want to try more translation models to optimize for domain adaptation. We will also explore the use of the audio signal to improve domain-specific information but as a post-processing step rather than optimizing an entire speech recognition system which may not be viable in all cases.